\documentclass[aps,twocolumn,pra,superscriptaddress,amsmath,showpacs,tightenlines,pdflatex]{revtex4}

\usepackage{amsmath}

\usepackage{graphicx,times}
\usepackage{amstext}
\usepackage{amsmath}            
\usepackage{amssymb}            
\usepackage{graphicx}           
\usepackage{latexsym}
\usepackage{bm}
 \usepackage{color}

\def \etal{\textit{et al.}}

\newcommand{\pla}{Phys. Lett. A~}

\def\tr{{\rm Tr}}

\def\CP{{\rm CP}}
\def\NP{{\rm NP}}

\def\<{\langle}
\def\>{\rangle}

\newcommand{\ket}[1]{\mbox{$|#1\rangle$}}
\newcommand{\bra}[1]{\mbox{$\langle#1|$}}

\begin{document}

\title{Statistical mixtures of states can be more quantum than their superpositions:\\
Comparison of nonclassicality measures for single-qubit states}

\author{Adam Miranowicz}
\affiliation{CEMS, RIKEN, 351-0198 Wako-shi, Japan}
\affiliation{Faculty of Physics, Adam Mickiewicz University,
61-614 Pozna\'n, Poland}

\author{Karol Bartkiewicz}
\affiliation{Faculty of Physics, Adam Mickiewicz University,
61-614 Pozna\'n, Poland}\affiliation{RCPTM, Joint Laboratory of
Optics of Palack\'y University and Institute of Physics of AS CR,
Palack\'y University, 17. listopadu 12, 771 46 Olomouc, Czech
Republic}

\author{Anirban Pathak}
\affiliation{Department of Physics and Materials Science and
Engineering, Jaypee Institute of Information Technology, A-10,
Sector-62, Noida, UP-201307, India}

\author{Jan Pe\v{r}ina Jr.}
\affiliation{RCPTM, Joint Laboratory of Optics of Palack\'y
University and Institute of Physics of AS CR, Palack\'y
University, 17. listopadu 12, 771 46 Olomouc, Czech Republic}

\author{Yueh-Nan Chen}
\affiliation{Department of Physics and National Center for
Theoretical Sciences, National Cheng-Kung University, Tainan 701,
Taiwan} \affiliation{CEMS, RIKEN, 351-0198 Wako-shi, Japan}

\author{Franco Nori}
\affiliation{CEMS, RIKEN, 351-0198 Wako-shi, Japan}
\affiliation{Department of Physics, The University of Michigan,
Ann Arbor, Michigan 48109-1040, USA}

\date{\today}

\begin{abstract}
A bosonic state is commonly considered nonclassical (or quantum)
if its Glauber-Sudarshan $P$ function is not a classical
probability density, which implies that only coherent states and
their statistical mixtures are classical. We quantify the
nonclassicality of a single qubit, defined by the vacuum and
single-photon states, by applying the following four well-known
measures of nonclassicality: (1) the nonclassical depth, $\tau$,
related to the minimal amount of Gaussian noise which changes a
nonpositive $P$ function into a positive one; (2)~the nonclassical
distance $D$, defined as the Bures distance of a given state to
the closest classical state, which is the vacuum for the
single-qubit Hilbert space; together with (3) the negativity
potential (NP) and (4) concurrence potential, which are the
nonclassicality measures corresponding to the entanglement
measures (i.e., the negativity and concurrence, respectively) for
the state generated by mixing a single-qubit state with the vacuum
on a balanced beam splitter. We show that complete statistical
mixtures of the vacuum and single-photon states are the most
nonclassical single-qubit states regarding the distance $D$ for a
fixed value of both the depth $\tau$ and NP in the whole range
$[0,1]$ of their values, as well as the NP for a given value of
$\tau$ such that $\tau>0.3154$. Conversely, pure states are the
most nonclassical single-qubit states with respect to $\tau$ for a
given $D$, NP versus $D$, and $\tau$ versus NP. We also show the
``relativity'' of these nonclassicality measures by comparing
pairs of single-qubit states: if a state is less nonclassical than
another state according to some measure then it might be more
nonclassical according to another measure. Moreover, we find that
the concurrence potential is equal to the nonclassical distance
for single-qubit states. This implies an operational
interpretation of the nonclassical distance as the potential for
the entanglement of formation.
\end{abstract}



\maketitle

\section{Introduction}

One of the central problems of quantum theory, already raised by
its founders~\cite{Einstein05,Neumann32,Einstein35,Schrodinger35},
is the question of testing whether a given physical system cannot
be properly described classically. This problem has attracted
special interest in quantum optics~\cite{DodonovBook,PerinaBook},
quantum information~\cite{NielsenBook,Horodecki09review}, and
recently even in quantum biology~\cite{Lambert13,Li12}. In this
paper, we address the problem of not only testing but also
quantifying nonclassicality (or quantumness) of light or, more
generally, of a bosonic system.

In general, a state is referred to as \emph{nonclassical} if its
Glauber-Sudarshan $P$~function~\cite{Glauber63,Sudarshan63} cannot
be considered a classical probability density~\cite{VogelBook},
which means that it is not positive (semidefinite). In other
words, a state that cannot be expressed as a statistical mixture
of coherent states is called nonclassical. Otherwise the state is
considered classical. It is worth noting that if the $P$~function
is more singular than the Dirac $\delta$-function (which is the
case for, e.g., the Fock states), then it is also nonpositive.
Thus, the nonpositivity of the $P$~function is the necessary and
sufficient condition for nonclassicality.

There exist several criteria of nonclassicality. However, most of
these criteria can only show a signature of nonclassicality. They
do not provide any quantitative measure of the nonclassicality.
Thus, they cannot be used to compare the amount of nonclassicality
present in two different states. Besides the above-mentioned
$P$-function-based criterion of nonclassicality, all the finite
sets of other criteria are sufficient but not necessary. Only an
infinite set (or hierarchy) of nonclassicality criteria can be
considered a sufficient and necessary condition of nonclassicality
(see, e.g., Ref.~\cite{Richter02}). Thus, these finite-set
criteria of nonclassicality may be better viewed as
\emph{witnesses} of nonclassicality rather than measures of
nonclassicality. This limitation of the existing criteria of
nonclassicality is well known and several efforts have been made
to quantify nonclassicality. These efforts led to the introduction
of various measures of nonclassicality.

For example, in 1987, Hillery introduced a distance-based measure
of nonclassicality~\cite{Hillery87}. Specifically, the trace
distance of a quantum state from the nearest classical state can
be considered as a measure of nonclassicality associated with the
given quantum state. This idea of distance-based measures has
attracted considerable attention in quantum
optics~\cite{Dodonov00,Wunsche01,Marian02,Marian04,Dodonov03}.
This intuitive definition is easy to understand but extremely
difficult to compute as it requires minimization over an infinite
number of variables. Specifically, one needs to minimize over the
set of all possible classical states in order to identify an
optimal reference classical state that yields a minimum distance
with respect to a given nonclassical state. This is the main
problem associated with the distance-based measures of
nonclassicality. Because of this computational difficulty, until
now the nonclassical distance has not exactly been computed for
any nonclassical state according to the original definition.
However, the computational difficulty associated with Hillery's
original measure can be circumvented by measuring the distance of
a given nonclassical state from a specific class of classical
states. This approach was adopted in a few works. For example,
Marian \etal~\cite{Marian02} defined a simplified version of the
Hillery nonclassical distance for a single-mode Gaussian state of
a radiation field as the Bures distance between the state and the
set of all classical single-mode Gaussian states. W\"unsche
\etal~\cite{Wunsche01,Dodonov00} measured the distance of a given
state from the set of only coherent states. Specifically, they
used the Hilbert-Schmidt distance of a pure state $\rho$ from the
coherent states as a quantitative measure of nonclassicality of
$\rho$~\cite{Wunsche01}. Almost in the similar line, Mari
\etal~\cite{Mari11} introduced a measure of nonclassicality of a
state $\rho$ in terms of its trace-norm distance from the set of
all states having the positive Wigner function. Strictly, speaking
this quantifier of nonclassicality is not a proper measure since
some nonclassical states do have positive Wigner function (as
discussed below with respect to a nonclassicality volume).
Similarly, Dodonov and Ren\'o~\cite{Dodonov03} used the
Hilbert-Schmidt distance from the set of all displaced thermal
states as a quantitative measure of nonclassicality. These
measures are naturally free from the problem that arises due to
the minimization over the set of arbitrary classical states.

In 1991, Lee~\cite{Lee91} introduced a quantitative measure of
nonclassicality which is usually referred to as nonclassical
depth. It is well known that noise can destroy nonclassicality.
Lee used this property to define the nonclassical depth as the
minimum amount of noise required to destroy the nonclassicality.
This measure is not continuous and for every non-Gaussian pure
state it is always equal to $1$~\cite{Lutkenhaus95}. As a
consequence, one cannot use this measure to compare the amount of
nonclassicality present in two non-Gaussian pure states. The
nonclassical depth was applied in dozens of papers (see, e.g.,
Refs.~\cite{Lee92,Lutkenhaus95,Malbouisson03} and for reviews see
Refs.~\cite{DodonovBook,Klyshko96}).

In 2004, Kenfack and \.Zyczkowski~\cite{Kenfack04} introduced the
concept of the nonclassical volume, which is a quantitative
parameter of nonclassicality corresponding to the volume of the
negative part of the Wigner function. A non-zero value of the
volume definitely indicates the existence of nonclassical state,
but this volume is not useful as a measure in general, since the
Wigner function cannot detect the presence of nonclassicality in
all quantum states. Specifically, the Wigner function of a
squeezed coherent state is not negative. As a consequence, the
nonclassical volume vanishes for all squeezed coherent states,
although they are nonclassical according to the definition based
on the nonpositivity of the $P$~function. This example implies
that, in general, the nonclassical volume is not an appropriate
measure of nonclassicality.

Various other methods to test (or witness) nonclassicality (see,
Ref.~\cite{VogelBook} and references therein) and quantify
it~\cite{Gehrke12,Vogel14,Mraz14} have been developed by Vogel
\etal~In particular, the nonclassicality
witnesses~\cite{Richter02}, based on the matrices of the
normally-ordered moments of, e.g., annihilation and creation
operators, have attracted considerable interest as an infinite set
of observable conditions corresponding to a necessary and
sufficient condition for nonclassicality. Various generalizations
have been studied, including tests of spatiotemporal nonclassical
properties of multimode fields~\cite{Vogel08,Miran10,
Bartkowiak11}. Moreover, this approach was the inspiration to
introduce entanglement witnesses based on the matrices of moments
of annihilation and creation operators of the partially-transposed
density matrices~\cite{Shchukin05,Miran06} (for generalizations
see, e.g., Refs.~\cite{Haseler08,Miran09}). The relations between
these entanglement and nonclassicality criteria were also studied
in detail (see, e.g., Ref.~\cite{Miran10}). Note that the majority
of these works have solely described nonclassicality (or
entanglement) witnesses rather than nonclassicality measures. Only
the more recent works of Vogel \etal~(see, e.g.,
Refs.~\cite{Gehrke12,Vogel14,Mraz14}) were focused on quantifying
nonclassicality. For example, an experimentally-accessible method
to determine a degree of nonclassicality was recently described in
Ref.~\cite{Mraz14}.

With the advances in quantum computation and information, many
measures of entanglement (which is a specific manifestation of
nonclassicality) have been studied. Unfortunately, measures of
entanglement cannot be applied directly to all nonclassical
states. For example, nonclassicality of single-mode states cannot
be measured directly by using a measure of entanglement.
Interestingly, an indirect way to use measures of entanglement as
measures of nonclassicality was suggested by Asboth
\etal~\cite{Asboth05}. Specifically, if a single-mode nonclassical
(classical) state is combined with the vacuum at a beam splitter,
then the output state will be entangled (separable), for which
various entangled measures can be applied. For example, in the
original Ref.~\cite{Asboth05}, the relative entropy of
entanglement and the logarithmic negativity (referred to as
entanglement potentials) were applied as measures of entanglement
produced at the output of a balanced beam splitter as the result
of combining a nonclassical state with the vacuum. In principle,
one can use any other measure of entanglement (e.g., the
concurrence related to the entanglement of formation) to measure
nonclassicality using this approach. Recently, Vogel and
Sperling~\cite{Vogel14} studied the approach in
Ref.~\cite{Asboth05} to measure nonclassicality based on the
Schmidt rank as an entanglement potential. Note that this measure
based on the Schmidt rank is discontinuous (analogously to the
nonclassical depth, as it is explained in detail in Sec. III~A).
Here we apply the continuous entanglement potentials, which are
based on the negativity and concurrence.

It is important to clarify our usage of the term
\emph{entanglement potential}, which is more general than that
used in the original Refs.~\cite{Asboth05,Vogel14}. Specifically,
we use this notion by referring to any entanglement measure
applied to the output of the auxiliary beam splitter used in
Ref.~\cite{Asboth05}. Thus, in our understanding, the following
measures can be considered as special cases of entanglement
potentials: the negativity and concurrence potentials, as well as
those based on (i) the logarithmic negativity, (ii) relative
entropy of entanglement, and (iii) Schmidt numbers. However,
strictly speaking, Asboth \etal~\cite{Asboth05} referred solely to
the measure (i) as the entanglement potential, while to the
measure (ii) as the entropic entanglement potential. Moreover,
Vogel and Sperling~\cite{Vogel14} are not referring to the measure
(iii) as an entanglement potential at all.

We analyze the nonclassicality of states only. Note that the
nonclassicality of operations (see, e.g.,
Refs.~\cite{Meznaric13,Lambert10,Emary14}) can also be studied by
applying various measures.

The discussion above shows that there exists a large number of
quantitative measures of nonclassicality. However, none of the
measures can be considered as superior as all of them have some
limitations and different physical (or operational)
interpretations. Here we discuss the relativity of a set of
nonclassicality measures which can be observed even for the
simplest nontrivial case of a single qubit defined as a coherent
or incoherent superposition of the vacuum and single-photon
states. We also report our analytical solutions for the Lee
nonclassical depth, the negativity potential, and the Hillery
nonclassical distance. The latter is found to be equivalent to the
concurrence potential. Further, we find boundary states, which are
maximally nonclassical states according to one nonclassicality
measure for a given value of another nonclassicality measure.

It is well known, and already confirmed
experimentally~\cite{Lvovsky02}, that statistical mixtures of the
vacuum and single-photon states are nonclassical (except for the
vacuum). We find, which is the most important result of this
paper, that such statistical mixtures can be more nonclassical
than coherent or partially incoherent superpositions of the vacuum
and single-photon states. This can be noticed by comparing their
nonclassicality for two chosen measures.

For the clarity of our presentation, we analyze the
algebraically-simplest nonclassical states, i.e., single-qubit
states, which can be written in a general form in the Fock basis
as follows:
\begin{equation}
\rho(p,x) \equiv
[\rho_{mn}] = \left[\begin{array}{cc}
1-p&x\\
x^*&p
\end{array}\right],
 \label{rho}
\end{equation}
where the parameters are $p\in[0,1]$, $|x|\in[0,\sqrt{p(1-p)}]$,
and $m,n=0,1$.

The paper is organized as follows. In Sec.~II, we recall the
definitions of four popular nonclassicality measures. And, more
importantly, we find analytical formulas for these measures for
arbitrary single-qubit states. In Sec.~III, we present the main
results of this paper, which show the relativity of ordering
states with respect to their degree of nonclassicality. We also
demonstrate that the nonclassicality of mixed states can exceed
that of superposition states. We conclude in Sec.~IV.

\section{Nonclassicality measures for single-qubit states}

\subsection{Nonclassical depth}

Here recall the concept of the nonclassical depth $\tau$
introduced by Lee~\cite{Lee91,Lee92} (for a review see
Ref.~\cite{DodonovBook} and references therein). We present the
definition of $\tau$ in a slightly different form as based on the
standard Cahill-Glauber $s$-parametrized quasiprobability
distribution (QPD) rather than the $R$-function used by Lee. Then
we find a compact formula for the nonclassical depth for arbitrary
single-qubit states.

We start from the Fock-state representation of the
$s$-parametrized QPD, $W^{(s)}(\alpha)$, for an
arbitrary-dimensional state $\rho$ as~\cite{Cahill69}
\begin{equation}
  W^{(s)}(\alpha)=\sum_{m,n=0}^\infty\rho_{mn}\bra{n}T^{(s)}(\alpha)\ket{m} ,
  \label{QPD1}
\end{equation}
given in terms of
\begin{equation}
  \bra{n}T^{(s)}(\alpha)\ket{m} = c
  \sqrt{\frac{n!}{m!}}y^{m-n+1}z^n(\alpha^*)^{m-n}L^{m-n}_n(x_{\alpha}),
\label{QPD2}
\end{equation}
where $s\in[-1,1]$, $c=\frac{1}{\pi}\exp[-2|\alpha|^2/(1-s)]$,
$x_{\alpha}=4|\alpha|^2/(1-s^2)$, $y=2/(1-s)$, $z=(s+1)/(s-1)$,
and $L^{k}_n(x_{\alpha})$ are the associate Laguerre
polynomials~\cite{SpanierBook}. In special cases,
$L^{k}_0(x_{\alpha})=1$ and $L^{k}_1(x_{\alpha})=1+k-x_{\alpha}$.
Moreover, $\alpha$ is a complex number, where its real and
imaginary parts can be interpreted as canonical position and
momentum, respectively. The operator $T^{(s)}(\alpha)$ is defined
in the Fock representation by Eq.~(\ref{QPD2}) or, equivalently,
by the formula
$T^{(s)}(\alpha)=\frac{1}{\pi}yz^{(a^{\dagger}-\alpha^*)(a-\alpha)}$, where $a$
($a^{\dagger}$) is the annihilation (creation) operator. In the
special cases of $s=-1,0,1$, the QPD $ W^{(s)}(\alpha)$ becomes
the Husimi $Q$, Wigner $W$, and Glauber-Sudarshan $P$~functions,
respectively.

For a general single-qubit state, Eq.~(\ref{QPD1}) reduces to
\begin{equation}
  W^{(s)}(\alpha)=c y[\rho_{00}+z(1-x_\alpha)\rho_{11}+2y{\rm Re}(\alpha \rho_{01})].
  \label{QPD3}
\end{equation}
As already explained, the standard definition of nonclassicality
is based on the nonpositivity of the $P$~function. The
$s$-parametrized QPDs can be used to quantify the degree of
nonclassicality. For example, the concept of the nonclassical
depth of Lee~\cite{Lee91} can be easily understood by recalling
the relation between two QPDs,  ${\cal W}^{(s_1)}$ and ${\cal
W}^{(s_2)}$ with $s_2<s_1$:
\begin{equation}
{\cal W}^{(s_2)} ( \alpha) =c'\int \exp\left( - \frac{2| \alpha -
\beta|^2}{s_1-s_2} \right) {\cal W}^{(s_1)} ( \beta) {\rm d}^2
\beta, \label{QPD4}
\end{equation}
where $c'=2/[\pi(s_1-s_2)]$. It is seen that all the QPDs can be
obtained from the $P$~function ($s_1=1$) by its convolution with
the Gaussian noise. By decreasing the parameter $s$ from 1, the
$P$~function for a given nonclassical state becomes non-negative
at some value (say $s_0$). This is because the Husimi function
($s=-1$) is non-negative for any state. The Lee nonclassical depth
$\tau$ is simply related to this Cahill-Glauber parameter $s_0$,
viz. $\tau=(1-s_0)/2$.

From the QPD, given by Eq.~(\ref{QPD3}), for a general
single-qubit state, we can write that
\begin{equation}
  \tau=\frac{1-s_0}{2} =\frac12-\frac12\min_{\alpha}s_-(\alpha),
 \label{tau1}
\end{equation}
where
\begin{equation}
  s_-(\alpha)=1+[2{\rm Re}(\alpha x)-p]-\sqrt{[2{\rm Re}(\alpha x)-p]^2-4p|\alpha|^2}.
 \label{depth}
\end{equation}
We found analytically the minimum of Eq.~(\ref{tau1}), which leads
to the following simple general formula for the nonclassical depth
of an arbitrary single-qubit state, given in Eq.~(\ref{rho}):
\begin{equation}
  \tau[\rho(p,x)] = \frac{\rho_{11}^2}{\rho_{11}-|\rho_{01}|^2} = \frac{p^2}{p-|x|^2},
 \label{tauSol}
\end{equation}
assuming $p\in(0,1]$ and $|x|\in[0,\sqrt{p(1-p)}]$. While for
$p=0$, the formula is simply given by $\tau[\rho(0,0)]=0$.

\subsection{Entanglement potentials}

Here we study the negativity and concurrence potentials as
measures of nonclassicality based on the unified description of
nonclassicality and entanglement by applying a beam-splitter (BS)
transformation as introduced in Ref.~\cite{Asboth05}.

The BS transformation can formally be described by the Hamiltonian
$H=\frac12(a^\dagger b+a b^\dagger)$, where
$a=\ket{0}\bra{1}=[0,1;0,0]$ and, analogously, $b$ are the
annihilation operators of the input modes. The unitary
transformation $U_{\rm BS}=\exp(-iHt)$ in the four-dimensional
Hilbert space can be written as
\begin{equation}
  U_{\rm BS} = \left[\begin{array}{cccc}
1&0&0&0\\
0&\cos(t/2)&-i\sin(t/2)&0\\
0&-i\sin(t/2)&\cos(t/2)&0\\
0&0&0&1\\
\end{array}\right],
\label{Ubs}
\end{equation}
where, for simplicity, we set $\hbar=1$. In general,
$T=\cos^2(t/2)$ and $R=\sin^2(t/2)$ correspond to the BS
transmittance and reflectance, respectively. A balanced beam
splitter (with $T=R$) corresponds to the evolution time $t=\pi/2$.

The state $\rho_{\rm out}$, which is the output of the BS with a
general single-qubit state $\rho$, given in Eq.~(\ref{rho}), and
the vacuum at the two input ports, is given by
\begin{equation}
  \rho_{\rm out} = U_{\rm BS} (\rho\otimes \ket{0}\bra{0} )U^\dagger_{\rm
  BS}.
\label{RhoOut1}
\end{equation}
In the special case for the balanced BS we have
\begin{equation}
  \rho_{\rm out}(p,x) = \left[\begin{array}{cccc}
1-p& \frac1{\sqrt{2}}ix &\frac1{\sqrt{2}}x& 0\\
-\frac1{\sqrt{2}}ix^*&\frac12 p&-\frac12ip&0\\
\frac1{\sqrt{2}}x^*& \frac12 i p&\frac12p&0\\
0&0&0&0\\
\end{array}\right].
\label{RhoOut2}
\end{equation}
The output state is entangled (except when the input is in the
vacuum state) as can be verified by applying entanglement
measures. Here we apply the negativity $N$ and concurrence $C$ for
the BS output state $\rho_{\rm out}$, which can be interpreted as
nonclassicality measures referred to as entanglement potentials of
an input state $\rho$.

\subsubsection{Negativity potential}

The negativity potential (NP) of a single-mode input state $\rho$
can be defined as the negativity $N$ of the two-mode output state
$\rho_{\rm out}$, i.e.,
\begin{eqnarray}
  \NP(\rho) &\equiv& N(\rho_{\rm out}). \label{NP}
\end{eqnarray}
Recall that the negativity for two qubits
is given by~\cite{Horodecki09review}
\begin{equation}
N({\rho_{\rm out}})=\max[0,-2\min{\rm eig}(\rho_{\rm out}^{\Gamma})],
\label{negativity}
\end{equation}
which is proportional to the negative eigenvalue of the
matrix $\rho_{\rm out}^{\Gamma}$ corresponding to the partial
transpose of $\rho_{\rm out}$ with respect to one of the
qubits. Thus, it is seen that the negativity corresponds to
the Peres-Horodecki separability condition based on the
partial transpose~\cite{Peres96,Horodecki96}. The negativity
[or more precisely the logarithmic negativity, $\log_2
(N+1)$] has an operational interpretation as the entanglement
cost under operations preserving the positivity of partial
transpose (PPT)~\cite{Audenaert03,Ishizaka04}. It was also
shown that the number of entangled degrees of freedom of two
subsystems can be estimated from the
negativity~\cite{Eltschka13}. Thus, in analogy to these
interpretations, the NP can be also referred to as the
entanglement potential for the estimation of entangled
dimensions or the potential for the PPT entanglement cost.

We find that the NP for an arbitrary single-qubit state
$\rho(p,x)$ can be given by the following formula:
\begin{equation}
  \NP[\rho(p,x)]=\frac{1}{3} \left[2 {\rm Re}\left(\sqrt[3]{2 \sqrt{a_1}+2 a_2}\right)+p-2\right],
 \label{NPgeneral}
\end{equation}
where
\begin{eqnarray}
a_{1} &=& a_2^2-2 [5 (p-1) p+6 |x|^2+2]^3,
\nonumber \\
a_{2} &=& 14 p^3-21 p^2+15 p+9 (p-2) |x|^2-4.
\label{NPgeneral2}
\end{eqnarray}
This solution was found by solving the following equation for the
negativity~\cite{Bartkiewicz15}:
\begin{eqnarray}
48\det\rho^\Gamma + 3N^4 + 6N^3 - 6N^2(\Pi_2 - 1) &&\nonumber\\
- 4N(3\Pi_2 - 2\Pi_3 - 1)  &=& 0, \label{Negativity2}
\end{eqnarray}
which is given in terms of the measurable and invariant moments
$\Pi_n = \tr[(\rho^\Gamma)^n]$. The negativity is given by a much
more complicated formula than those for any other nonclassicality
measures studied here. Surprisingly, a direct calculation of the
eigenvalues of $\rho^{\Gamma}_{\rm out}$ can result even in a more
complicated formula. Of course, Eq.~(\ref{NPgeneral}) can be
considerably simplified in special cases. For example, the NP for
$\rho'=\rho(p=1/8,x=1/4)$ reads
\begin{equation}
  \NP(\rho') =\tfrac14\sqrt{26}\cos\Big\{\tfrac13
  \Big[\pi-{\rm arctg}\Big(\tfrac{1}{66}\sqrt{38}\Big)\Big]
  \Big\}-\tfrac58.
\label{N1.33}
\end{equation}
The NP for other special states, which are important in our
comparisons, are analyzed in Sec.~III.

\subsubsection{Concurrence potential}

In analogy to the NP, the concurrence potential (CP) of a given
single-qubit state $\rho$ can be given in terms of the concurrence
$C$ of the two-qubit output state $\rho_{\rm out}$, viz.,
\begin{eqnarray}
  \CP(\rho) &\equiv& C(\rho_{\rm out}). \label{CP}
\end{eqnarray}
The concurrence for a general two-qubit system is defined
as~\cite{Wootters98}:
\begin{equation}
C({\rho_{\rm out}})=\max \Big\{0,2\max_j\lambda_j-\sum_j\lambda_j\Big\},
\end{equation}
where $\{\lambda^2 _{j}\} = \mathrm{eig}[{\rho }_{\rm out}({\sigma
}_{2}\otimes {\sigma }_{2}){\rho}_{\rm out}^{\ast }({ \sigma }_{2}\otimes
{\sigma }_{2})]$, and ${\sigma }_{2}$ is the Pauli operator. This
measure is monotonically related to the entanglement of formation,
$E_{F}$, as follows~\cite{Wootters98}:
\begin{equation}
E_{F}=
h\left(\textstyle{\frac{1}{2}}[1+\sqrt{1-C^2}]\right),
\label{EoF}
\end{equation}
which is given via the binary entropy $h(x)=-x\log_2
x-(1-x)\log_2(1-x)$. Thus, the CP can also be referred to as the
potential for the entanglement of formation. A direct calculation
of the CP of $\rho(p,x)$ leads us to a particularly simple formula
\begin{equation}
  \CP[\rho(p,x)]=1-\<00|\rho_{\rm out}|00\>=\rho_{11}=p,
 \label{conc}
\end{equation}
for $p\in[0,1]$ and $|x|\in[0,\sqrt{p(1-p)}]$.

\subsection{Nonclassical distance}

Here we calculate the nonclassical distance $D$, which is the
Hillery measure of nonclassicality (see, for a review,
Ref.~\cite{DodonovBook} and references therein) for a specifically
chosen set of classical states. We also show that this distance is
equivalent to the CP for single-qubit states.

The nonclassical distance $D$  of a state $\rho$ can be defined as
the distance of $\rho$ to the nearest state from the set of all
classical states ${\cal C}$ as~\cite{Hillery87,Marian02}:
\begin{equation}
D(\rho)=\frac{1}{2}\min_{\sigma\in {\cal C}}{\cal
D}_{\mathrm{B}}^{2}(\rho,\sigma).
\end{equation}
In this paper, and contrary to the original
Refs.~\cite{Hillery87,Hillery89}, we assume the distance to be the
Bures metric $D_{\mathrm{B}}(\rho,\sigma)$~\cite{Bures69}, or
equivalently the Helstrom metric~\cite{Helstrom67}, which is
simply related to the fidelity $F(\rho,\sigma)$ as follows
\begin{equation}
{\cal D}^2_{\mathrm{B}}(\rho,\sigma)=2[1-\sqrt{F(\rho,\sigma)}]\,.
\end{equation}
The fidelity is defined as~\cite{BengtssonBook}
\begin{eqnarray}
  F(\rho,\sigma)=\Big(\tr\sqrt{\sqrt{\sigma}\rho\sqrt{\sigma}}\Big)^2,
\label{fidelity1}
\end{eqnarray}
which can also be interpreted as a transition
probability~\cite{Uhlmann76}, or a quantum generalization of the
Fisher information metric. Several methods for measuring or
estimating the fidelity are known (see
Ref.~\cite{Bartkiewicz13fidelity} and references therein). The
fidelity for single-qubit states simplifies to
\begin{eqnarray}
  F(\rho,\sigma)=\tr(\rho\sigma)+\sqrt{(1-\tr\rho^2)(1-\tr\sigma^2)}.
\label{fidelity2}
\end{eqnarray}
We mention that the Bures distance can be applied in quantifying
not solely nonclassicality~\cite{Marian02}. It has also found
applications as indicators or measures of, e.g., state
distinguishability~\cite{Braunstein94}, quantum
entanglement~\cite{Vedral97,Marian08}, quantum
criticality~\cite{Zanardi08}, and light
polarization~\cite{Klimov05}.

It should be stressed that we look for the classical (or, least
nonclassical) states, belonging to the Hilbert space of an
investigated finite-dimensional system. Here we analyze the
Hilbert space of a single qubit, defined as a superposition of the
vacuum and single-photon Fock state. In this case the only
classical state is the vacuum. Thus, we set
$\sigma=\ket{0}\bra{0}$, then $F(\rho,\ket{0})=\rho_{00}=1-p$.
Thus,  it is seen that such defined nonclassical distance is
exactly equal to the CP,
\begin{equation}
  D[\rho(p,x)] = \CP[\rho(p,x)]=p,
 \label{DSol}
\end{equation}
for any values of $p\in[0,1]$ and $|x|\in[0,\sqrt{p(1-p)}]$. This
correspondence provides another quantum information interpretation
of the nonclassical distance.

We emphasize again that a nonclassical distance can be
defined differently, both by choosing another distance
measure and by extending the class ${\cal C}$ of classical
states, for which the minimization is performed. For example,
in the original papers of Hillery~\cite{Hillery87,Hillery89},
the trace norm was used as a distance measure. While Dodonov
\etal~\cite{Dodonov00,Wunsche01,Dodonov03} applied the
Hilbert-Schmidt distance. Moreover, the Kullback-Leibler
distance~\cite{Kullback59}, which is also known as
information divergence, information gain, or relative
entropy, can also be applied for quantifying nonclassicality,
in analogy to the entanglement measures based on the relative
entropy of
entanglement~\cite{Vedral97,Vedral98,Miran04b,Miran08a,Miran08b,Horst13}.

\begin{table}[bp] 
\caption{Examples of states satisfying all four special cases of
the inequalities given in Eq.~(\ref{ineq1}), where
$\rho_0=\rho(\frac12,\frac14)$, $\rho_{\rm P}(p)\equiv
\rho[p,\sqrt{p(1-p)}]$ is a pure state, and $\rho_{\rm M}(p)\equiv
\rho(p,0)$ is a completely mixed state.}
\begin{tabular}{l l l}
\hline
Case \quad\quad\quad   & Nonclassicality measures \quad\quad \quad   &  Examples of states $\rho$ \\
\hline
1 & $\tau(\rho)=D(\rho)=\NP(\rho)$ &  $\ket{0}, \ket{1}$\\
2 & $\tau(\rho)>D(\rho)>\NP(\rho)$ &  $\rho_0$\\
3 & $\tau(\rho)>D(\rho)=\NP(\rho)$ &  $\rho_{\rm P}(p)$ for $p\in(0,1)$\\
4 & $\tau(\rho)=D(\rho)>\NP(\rho)$ &  $\rho_{\rm M}(p)$ for $p\in(0,1)$\\
\hline
\end{tabular}
\end{table}
\section{Comparison of nonclassicality measures}

In general, for any single-qubit state $\rho=\rho(p,x)$, the
following inequalities hold
\begin{equation}
  \tau(\rho)\ge D(\rho) = \CP(\rho) \ge \NP(\rho).
 \label{ineq1}
\end{equation}
The left-hand inequality in Eq.~(\ref{ineq1}) can be deduced by
comparing explicitly the general expression for $\tau$ and $D$,
given by Eqs.~(\ref{tauSol}) and~(\ref{DSol}), respectively. The
right-hand inequality in Eq.~(\ref{ineq1}) is equivalent to the
well-known inequality $C(\rho')\ge N(\rho')$ for the concurrence
and negativity for arbitrary two-qubit states $\rho'$. Thus, in
particular, for the states $\rho_{\rm out}$ generated by the BS
from a single-qubit state $\rho$ and the vacuum. Table~I lists all
four special cases of these inequalities, together with examples
of states satisfying these cases.

In the following we analyze all the boundary states shown in
Figs.~1 and~2, and discuss the relativity of nonclassical measures
(see Tables~II and~III, and Fig.~3).

\begin{widetext}
\begin{figure*}
\centering
\includegraphics[scale=.37]{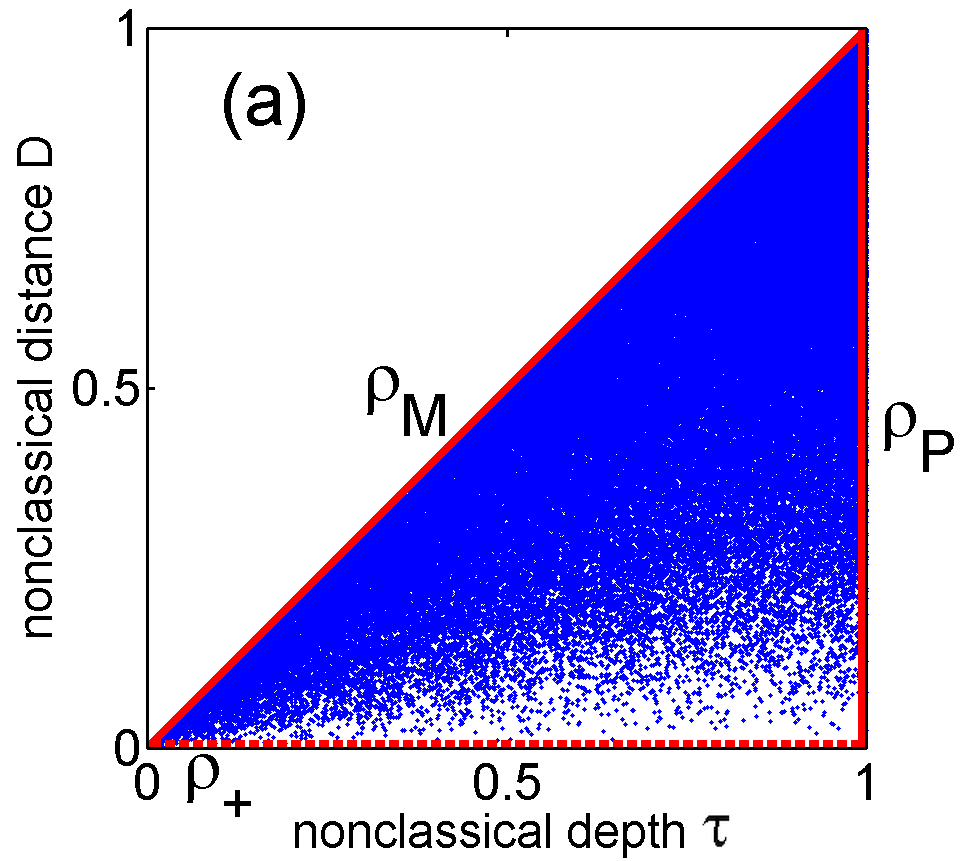}
\includegraphics[scale=.37]{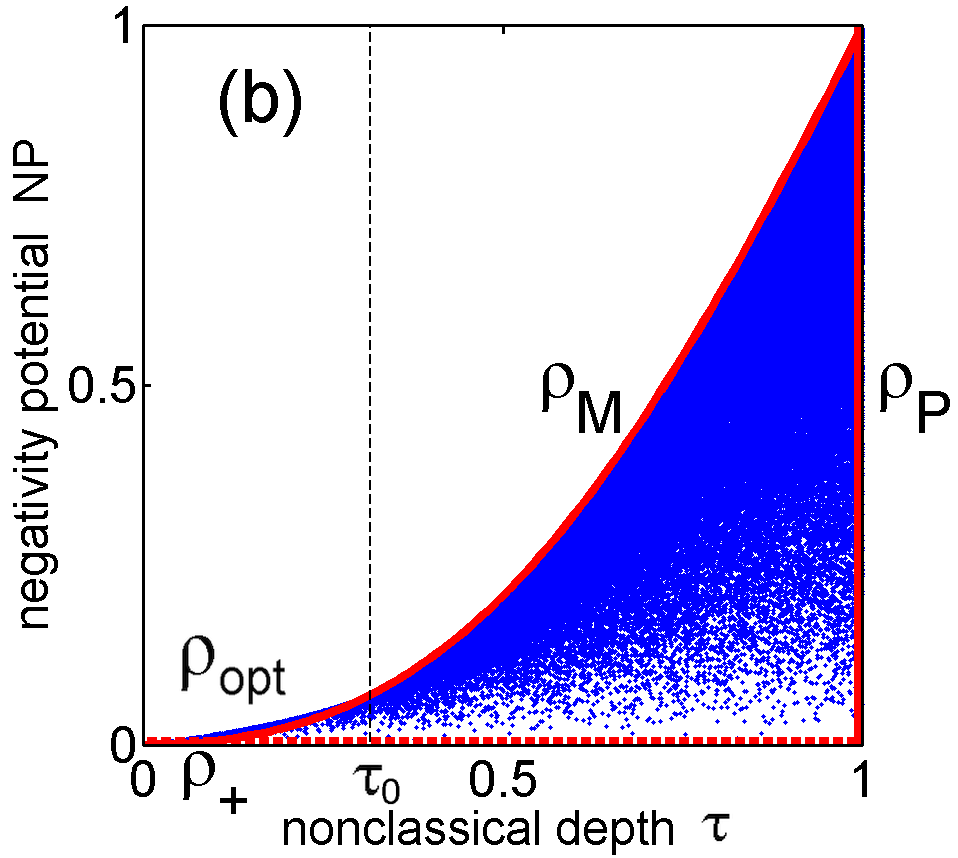}
\includegraphics[scale=.37]{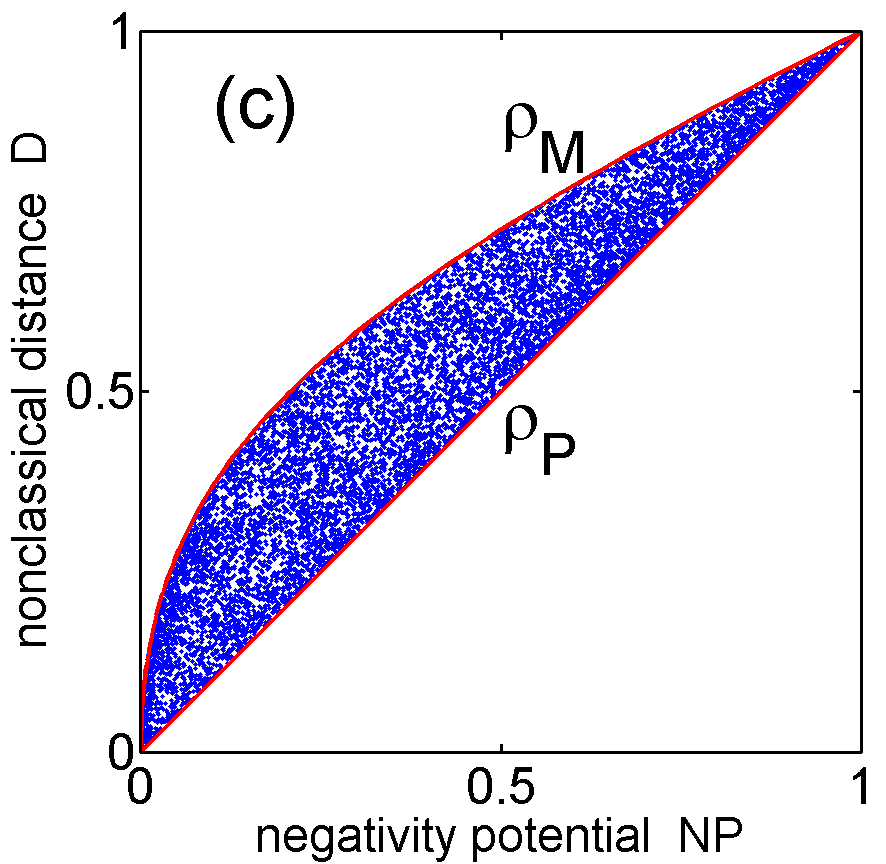}
 \caption{(Color online) Allowed values of the nonclassicality
measures for single-qubit states: (a) nonclassical distance $D$
versus nonclassical depth $\tau$, (b) negativity potential NP
versus $\tau$, and (c) $D$ versus NP. The points correspond to a
Monte Carlo simulation of $10^5$ states $\rho$. Each point is
plotted for $[D(\rho),\tau(\rho)]$ in (a) and analogously for
panels (b) and (c). The vertical broken line in panel (b) is
plotted at $\tau_0\approx 0.3154$. The boundaries are given by
pure states $\rho_{\rm P}$ [vertical red lines in the far right of
(a) and (b), and the red diagonal line in (c)], completely mixed
states $\rho_{\rm M}$ (solid red upper curves), as well as
partially mixed states $\rho_+$ (bottom broken lines) and
$\rho_{\rm opt}$ [corresponding to blue points right above the
curve for $\rho_{\rm M}$ in (b)]. In (b), it is barely visible
that $\rho_{\rm M}$ is \emph{not} the upper bound for
$\tau<\tau_0$. Thus, this region is magnified in Fig.~2(a). Note
that, in a mathematical sense, there are no states corresponding
exactly to the broken lines at the bottom of (a) and (b) for
$0<\tau\le 1$ and $D=\NP=0$. However, one can find states being
arbitrarily close to these lines.}
\end{figure*}
\end{widetext}

\begin{figure}
\includegraphics[scale=.45]{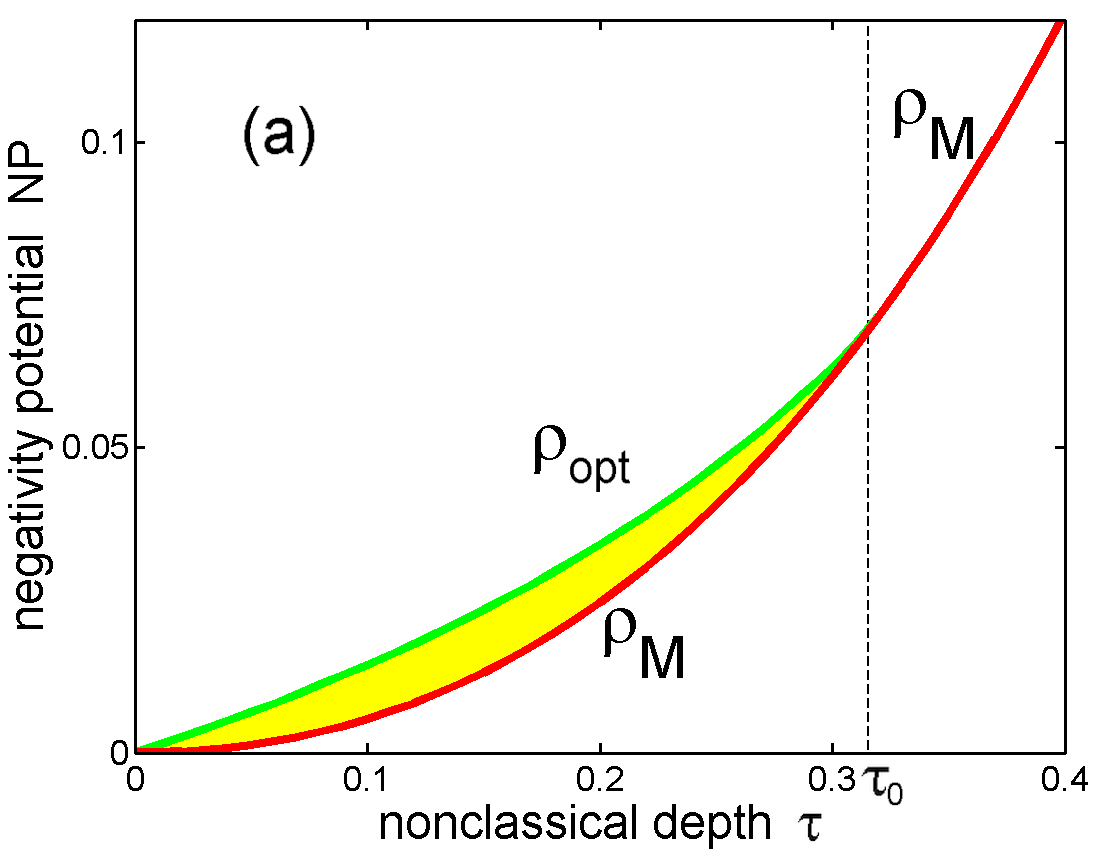}
\includegraphics[scale=.45]{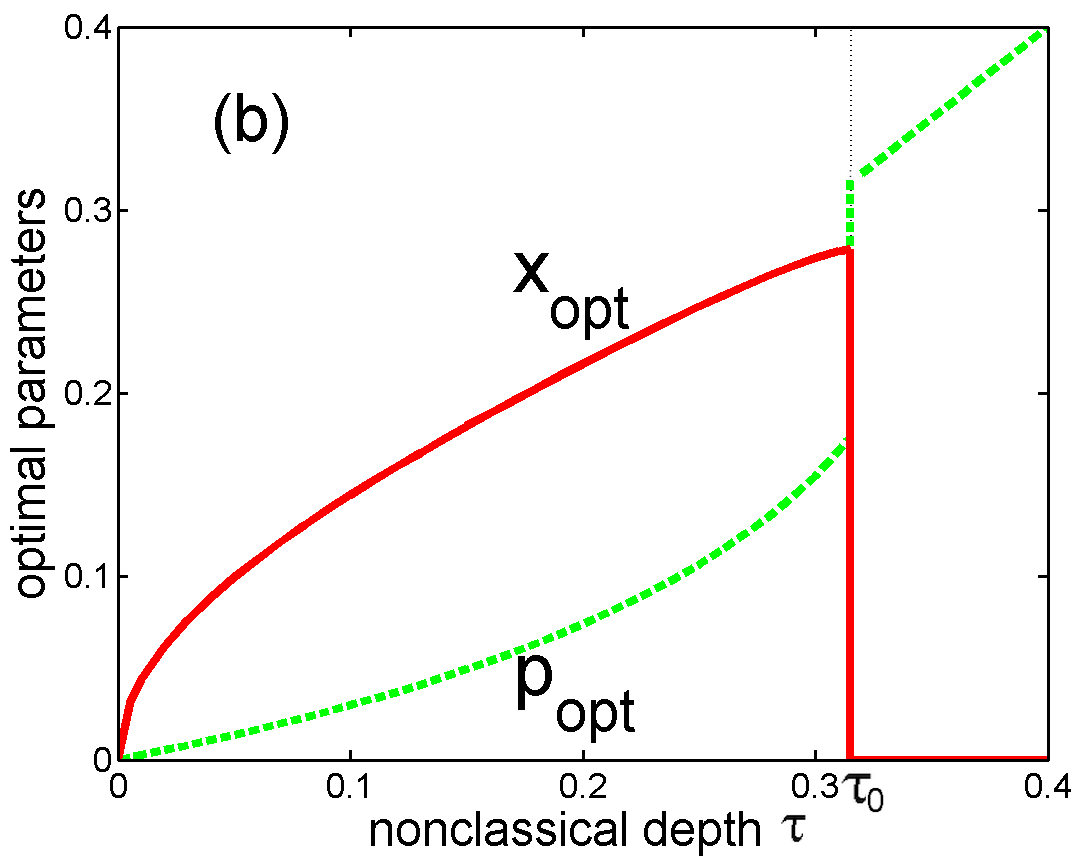}
 \caption{(Color online)
(a) The inset of Fig.~1(b) showing, in greater detail, the
boundaries for the NP versus nonclassical depth $\tau$. These
boundaries are reached by the partially mixed optimal states
$\rho_{\rm opt}$ for $\tau<\tau_0$ and completely mixed
states $\rho_{\rm M}$ for $\tau\ge \tau_0$. For clarity, we
do not plot here points corresponding to our Monte Carlo
simulation shown in Fig.~1. (b) Optimal parameters $p_{\rm
opt}=\langle 1|\rho_{\rm opt}|1\rangle$ and $x_{\rm
opt}=|\langle 0|\rho_{\rm opt}|1\rangle|$ as a function of
$\tau$. These parameters are discontinuous at $\tau=\tau_0$,
but, for clarity, we have plotted the red and green vertical
connecting lines at this point.}
\end{figure}

\subsection{Boundary states}

Figure~1 shows the nonclassicality regions for arbitrary
single-qubit states. The points $[X(\rho),Y(\rho)]$ in these
regions are obtained for the generated $10^5$ states $\rho$ by
performing a Monte Carlo simulation. Here  $X$ and $Y$ correspond
to chosen nonclassicality measures. Thus, by analyzing these
graphs in Fig.~1, one can say that some states are the most or
least nonclassical in terms of a measure $X$ \emph{for a given
value} of a measure $Y$.

Here we analyze the special cases of the general single-qubit
state $\rho$, which correspond to the single-qubit boundary states
shown in Figs.~1-3. We calculate the above-defined nonclassicality
measures for these states. In the Appendix, we present the proofs
that they are indeed the boundary states.

\subsubsection{Pure states}

Equation~(\ref{rho}) for $x=\sqrt{p(1-p)}$ reduces to a pure
state $\rho_{\rm P}=\ket{\psi_p}\bra{\psi_p},$ where
\begin{equation}
  \ket{\psi_p} = \sqrt{1-p}\ket{0}+\sqrt{p}\ket{1}.
 \label{psi_p}
\end{equation}
The BS output state for the input states $\ket{\psi_p}$ and
$\ket{0}$, is simply given by
\begin{equation}
|\psi_{\rm out} \rangle
=\sqrt{1-p}|00\>+\sqrt{\tfrac{p}{2}}(|10\>-i|01\>).
\label{PsiRhoOut}
\end{equation}
By recalling that
\begin{equation}
C(|\psi_{\rm out} \rangle)=N(|\psi_{\rm out} \rangle)
=2|c_{00}c_{11}-c_{01}c_{10}| \label{concP}
\end{equation}
for a general two-qubit pure state $|\psi \rangle
=\sum_{m,n=0,1}c_{mn}|mn\rangle$, where $c_{mn}$ are the
normalized complex amplitudes, one can obtain the nonclassical
measures as follows
\begin{equation}
  \NP(|\psi_{p} \rangle)=D(|\psi_{p}\rangle)=\rho_{11}=p.
\label{Npure}
\end{equation}
By contrast to these equal measures, the nonclassical depth for a
pure state reads
\begin{equation}
  \tau(|\psi_{p} \rangle)=1-\delta_{p,0},
 \label{TauPure}
\end{equation}
in terms the Kronecker delta $\delta_{p,0}$. In the special cases
of the vacuum and single-photon states, this formula reduces to
the known results~\cite{Lee92}. It is clearly seen that the depth
$\tau$ is discontinuous, as $\tau[\ket{\psi(p=1)}]\equiv
\tau(\ket{0})=0$, while $\tau[\ket{\psi(p>0)}]=1$, even for $p$
very close to zero. Note that also the entanglement potential
based on the Schmidt number is discontinuous.

Pure states are the boundary states in the three panels of Fig.~1
for the whole range $[0,1]$ of the ordinate. In particular, they
correspond to the lower bound of the nonclassical distance versus
NP. Note that we are analyzing the potential based on the
negativity rather than the logarithmic negativity, as suggested
and applied in Ref.~\cite{Asboth05}. Thus, the lower bound in
Fig.~1(c) is given by a straight line, which would not be the case
otherwise.

\subsubsection{Completely mixed states}

In another special case, Eq.~(\ref{rho}) for $x=0$ describes a
completely mixed state,
\begin{equation}
  \rho_{\rm M} = (1-p)\ket{0}\bra{0}+p\ket{1}\bra{1},
 \label{RhoM}
\end{equation}
i.e., a statistical mixture of the vacuum $\ket{0}$ and
single-photon state $\ket{1}$. Thus, we have
\begin{equation}
  \tau(\rho_{\rm M})=D(\rho_{\rm M})=p.
\label{TauMixed}
\end{equation}
The NP for any mixed state $\rho_{\rm M}(p)$ can be found from the
general formula given in Eq.~(\ref{NPgeneral}), but here we apply
a more explicit and intuitive derivation. Specifically, if the
input qubit state is completely mixed, then one finds that the BS
output state reads
\begin{eqnarray}
  \rho_{\rm out}(p,0) &=& U_{\rm BS} [\rho_{\rm M}(p)\otimes \ket{0}\bra{0} ]U^\dagger_{\rm BS}
\nonumber \\
   &=&  p\ket{\bar\psi^-}\bra{\bar\psi^-}+(1-p)\ket{00}\bra{00},
 \label{RhoOutM}
\end{eqnarray}
where $\ket{\bar\psi^-}=(\ket{10}-i\ket{01})/\sqrt{2}$. This is
the statistical mixture of a maximally entangled state and a
separable state orthogonal to it, which is often referred to as
the Horodecki state~\cite{Horodecki09review}. Such mixtures are
often studied in the comparisons of various entanglement and
nonlocality measures~\cite{Miran04a,Miran04b, Miran08b,Horst13,
Bartkiewicz13}. Thus, the NP for a mixed $\rho_{\rm M}(p)$ reads
as
\begin{equation}
  \NP(\rho_{\rm M}) = N[\rho_{\rm out}(p,0)]=\sqrt{(1-p)^2+p^2}-(1-p).
 \label{NP_RhoM}
\end{equation}
Completely mixed states are the boundary states shown in the three
panels of Fig.~1. However, it is worth noting that they are
\emph{not} extremal for the whole range of $\tau$ in Fig.~1(b),
which is shown in detail in Fig.~2(a) and discussed in the next
paragraph.

\subsubsection{Partially mixed optimal states}

A preliminary analysis of Fig.~1(b) can lead to a conjecture
that: completely mixed states $\rho_M$ correspond to the
upper boundary of the NP for an arbitrary value of the depth
$\tau\in[0,1]$. However, a closer scrutiny of Fig.~2(a),
which is the inset of Fig.~1(b), indicates that $\rho_M$ is
the extremal state only for $\tau\ge \tau_0$. This critical
value is $\tau_0\approx 0.3154$, as marked by the vertical
broken lines in Figs.~1(b) and~2. By contrast to this, there
are other states exhibiting higher nonclassicality if
$\tau<\tau_0$. Thus, let us define the following partially
mixed state
\begin{equation}
  \rho_{\rm opt}(\tau) \equiv \rho[p_{\rm opt}(\tau),x_{\rm
  opt}(\tau)],
\label{RhoOpt}
\end{equation}
where $x^2_{\rm opt}(\tau)=p_{\rm opt}(\tau)-p^2_{\rm
opt}(\tau)/\tau$, which corresponds to the maximum NP for a
given $\tau$ [as shown in Fig.~2(a)], i.e.,
\begin{equation}
  \NP(\rho_{\rm opt}(\tau))\equiv \max_p
  \NP[\rho(p,\sqrt{p-p^2\tau^{-1}})].
 \label{Nopt}
\end{equation}
The optimal matrix elements $p_{\rm opt}$ and $x_{\rm opt}$ are
shown as a function of $\tau$ in Fig.~2(b). These elements can
easily be obtained by numerically maximizing
Eq.~(\ref{NPgeneral}), with $|x|^2=p-p^2/\tau$, for a given
$\tau$. It is seen that $p_{\rm opt}=\tau$ and $x_{\rm opt}=0$ for
$\tau\ge \tau_0$, thus $\rho_{\rm opt}$ becomes  $\rho_M$ in this
range of $\tau$. Unfortunately, we have not found a compact-form
analytical expression for $\rho_{\rm opt}$ for $\tau<\tau_0$.

\subsubsection{Partially mixed states with nonzero $\tau$ for vanishing $D$ and NP}

We also analyze the state $\rho(p,x)$ defined in the right-hand
limit $p\rightarrow 0+$ with properly chosen $x$ as follows:
\begin{equation}
  \rho_+(\tau_{0})=\lim_{p\rightarrow0+} \rho(p,x_{0}),
 \label{RhoPlus}
\end{equation}
where
\begin{equation}
x_{0}=\sqrt{(1+p-\tau_0^{-1}p)p(1-p)},
 \label{x0}
\end{equation}
assuming $\tau_0\in(0,1]$. Note that pure states with
$\tau_0=1$ can also be considered here. To be more explicit, let us
analyze the special case of $\rho(p,x_{0})$, when $\tau_0=1/2$ and
\begin{equation}
p=D[\rho(p,x_{0})]=10^{-n} \Rightarrow \tau[\rho(p,x_{0})]=\frac{1}{2-10^{-n}},
 \label{x00}
\end{equation}
for $n=0,1,2,\ldots<\infty$. It is seen that the nonclassical
depth is approaching the chosen nonzero value $\tau_0=1/2$ at the
same rate as the nonclassical distance is vanishing. In general,
we can write:
\begin{eqnarray}
  \tau[\rho_{+}(\tau_0)]&\equiv& \lim_{p\rightarrow0+} \tau[\rho(p,x_{0})] = \tau_0,
\nonumber \\
  \NP[\rho_{+}(\tau_0)]&\equiv& \lim_{p\rightarrow0+} \NP[\rho(p,x_{0})] = 0,
\nonumber \\
  D[\rho_{+}(\tau_0)]&\equiv& \lim_{p\rightarrow0+} D[\rho(p,x_{0})] =
  0.
 \label{RhoPlusNoncl}
\end{eqnarray}
Thus, this state approaches the lower bound of the distance $D$
versus depth $\tau$ [shown as the bottom broken line in Fig.~1(a)]
and the NP versus depth $\tau$ [see bottom of Fig.~1(b)]. Because
of the discontinuity of the depth $\tau$, these lower bounds in
Figs.~1(a) and~1(b) are not exactly reached, as indicated by the
broken lines. This is also reflected in the definition of $\rho_+$
given by the right-hand limit in Eq.~(\ref{RhoPlus}). Thus,
strictly speaking $\NP(\rho)=0$ (or, equivalently, $D(\rho)=0$)
for a given state $\rho$ \emph{if and only if} $\tau(\rho)=0$.
This is because all these quantities are \emph{measures (rather
than only witnesses)} of nonclassicality, and thus they give the
necessary and sufficient conditions for the nonclassicality of an
arbitrary single-qubit state $\rho$.

\subsection{Mixtures of states can be more nonclassical than their superpositions}

The analysis of Fig.~1 can lead to the conclusion that the
completely mixed states $\rho_{\rm M}$ are the most nonclassical
single-qubit states with respect to: (i) the distance $D$ for a
given value of the depth $\tau\in[0,1]$, (ii) $D$ for a fixed
value of the $\NP\in[0,1]$, and (iii) the NP for a given value of
$\tau\in[\tau_0,1]$. Conversely, pure states $\rho_{\rm P}$ are
the most nonclassical single-qubit states regarding $\tau$ versus
$D$, $\tau$ versus NP, and NP versus~$D$.

This interpretation of the maximum nonclassicality of mixed states
should not be confused with the following conclusion that
dephasing could increase the nonclassicality. Such dephasing
results in decreasing the off-diagonal term $x$, while keeping the
diagonal terms unchanged. Specifically, one can observe that
\begin{eqnarray}
  \tau[\rho_{\rm P}(p)]\ge& \tau[\rho(p,x)] &\ge \tau[\rho_{\rm M}(p)],
\nonumber \\
  \NP[\rho_{\rm P}(p)]\ge& \NP[\rho(p,x)] &\ge \NP[\rho_{\rm M}(p)],
\nonumber \\
  D[\rho_{\rm P}(p)]=& D[\rho(p,x)] &= D[\rho_{\rm M}(p)],
 \label{ineq2}
\end{eqnarray}
for any $x\in[0,1]$. It is seen that by decreasing $x$, also
$\tau[\rho(p,x)]$ and $\NP[\rho(p,x)]$ decrease, while only
$D[\rho(p,x)]$ remains unchanged. Thus, in this interpretation
based on the inequalities in Eq.~(\ref{ineq2}), a mixed state
$\rho_{\rm M}(p)$ is not more nonclassical than a pure state
$\rho_{\rm P}(p)$ assuming the same element~$p$.

Our reverse conclusion about \emph{mixed states, which are
more nonclassical than superposition states} (including pure
states), refers to another comparison. To show this more
explicitly, we express $\rho(p,x)$ in terms of some
nonclassicality measures instead of the parameters $p,x$. In
particular, by inverting Eq.~(\ref{tauSol}) for
$\tau=\tau[\rho(p,x)]$ and by applying $D=D[\rho(p,x)]=p$,
one can express a general single-qubit state (assuming real
$x$) in terms of these nonclassicality measures, i.e.,
\begin{equation}
\rho(p,x) \equiv \rho'(D,\tau) = \left[\begin{array}{cc}
1-D&\sqrt{D-D^2\tau^{-1}}\\
\sqrt{D-D^2\tau^{-1}}&D
\end{array}\right],
 \label{rhoDtau}
\end{equation}
where $\tau\in[0,1]$ and $D\in[0,\tau]$. Analogously, we can
express $\rho(p,x)$ in terms of other pairs of nonclassicality
measures, e.g.,
\begin{eqnarray}
\rho(p,x) \equiv \rho''(N,\tau) = \rho'''(N,D),
 \label{rhoNtau}
\end{eqnarray}
where $N=\NP[\rho(p,x)]$, although the expressions will be much
more complicated here. Analogously, we introduce the symbols
$\rho'_{\rm M}$, $\rho''_{\rm M}$, and $\rho_{\rm M}'''$, denoting
the mixed state $\rho_{\rm M}$, which is expressed via the
nonclassical measures analogously to $\rho'$, $\rho''$, and
$\rho'''$, respectively. Note that the assumption of real $x$
follows from the property that the nonclassical measures $\tau$
and NP depend solely on the absolute value of $x$, while $D$ is
completely independent of $x$.

Thus, for a given value of the nonclassical depth, say
$\tau_1\in[0,1]$, one can observe that
\begin{eqnarray}
  D[\rho'_{\rm M}(D_1,\tau_1)] &\ge& D[\rho'(D',\tau_1)],
\label{ineq3a}
\end{eqnarray}
where $D'\in[0,\tau_1]$ and $D_1=\tau_1$. For a given value of the
depth $\tau_1\in[\tau_0,1]$, where $\tau_0=0.3154$, one finds that
\begin{eqnarray}
  \NP[\rho''_{\rm M}(N_1,\tau_1)] &\ge& \NP[\rho''(N'',\tau_1)],
\label{ineq3b}
\end{eqnarray}
where $N''\in[N_0,N_1]$ and $N_i=\sqrt{(1-\tau_i)^2+\tau_i^2}
-(1-\tau_i)$ for $i=0,1$. Moreover, for a given value of the
NP, say $N_1\in[0,1]$, one observes that
\begin{equation}
  D[\rho'''_{\rm M}(D_1,N_1)] \ge D[\rho'''(D''',N_1)],
\label{ineq3c}
\end{equation}
where $D'''\in[N_1,1]$ and here $D_1=\sqrt{2N_1(1+N_1)}-N_1$. All
these three inequalities show that completely mixed states can be
considered as the most nonclassical single-qubit states for a
fixed value of a proper nonclassical measure, as shown in the
corresponding panels of Fig.~1.

\begin{widetext}
\begin{figure*}
\includegraphics[scale=.38]{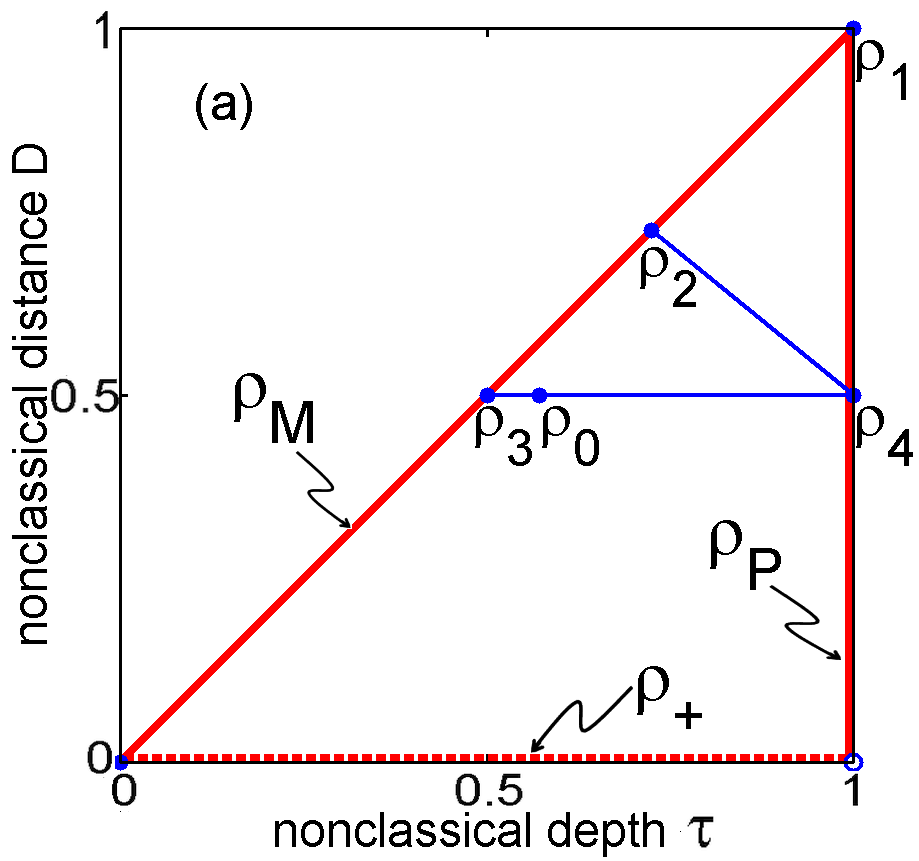}
\includegraphics[scale=.38]{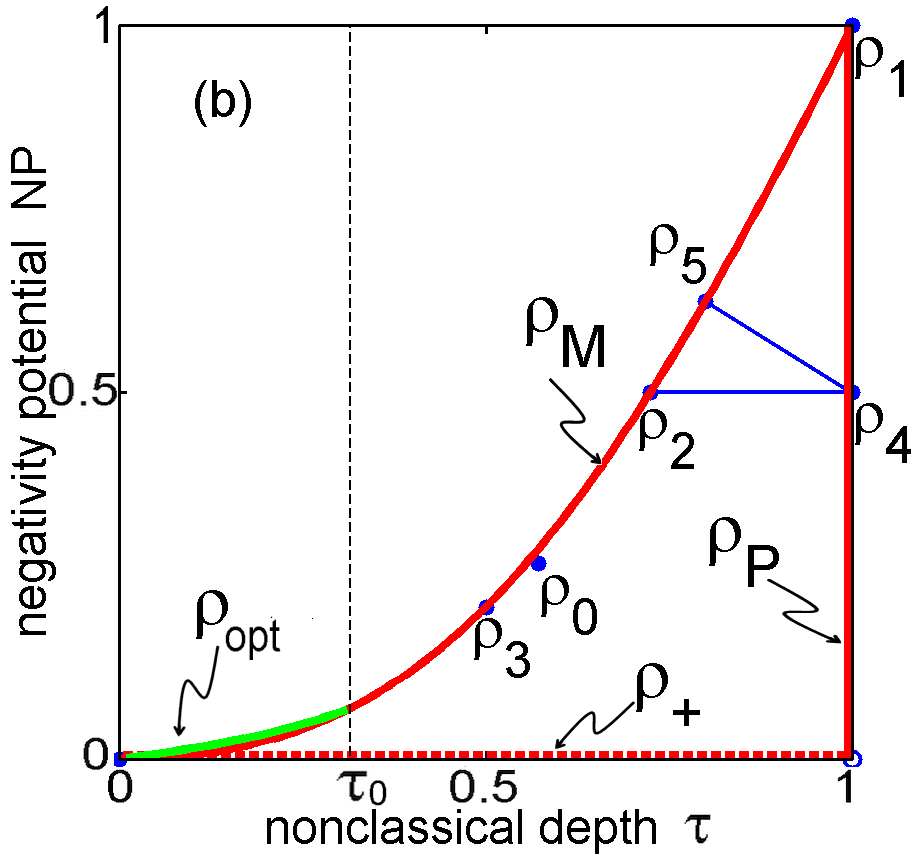}
\includegraphics[scale=.38]{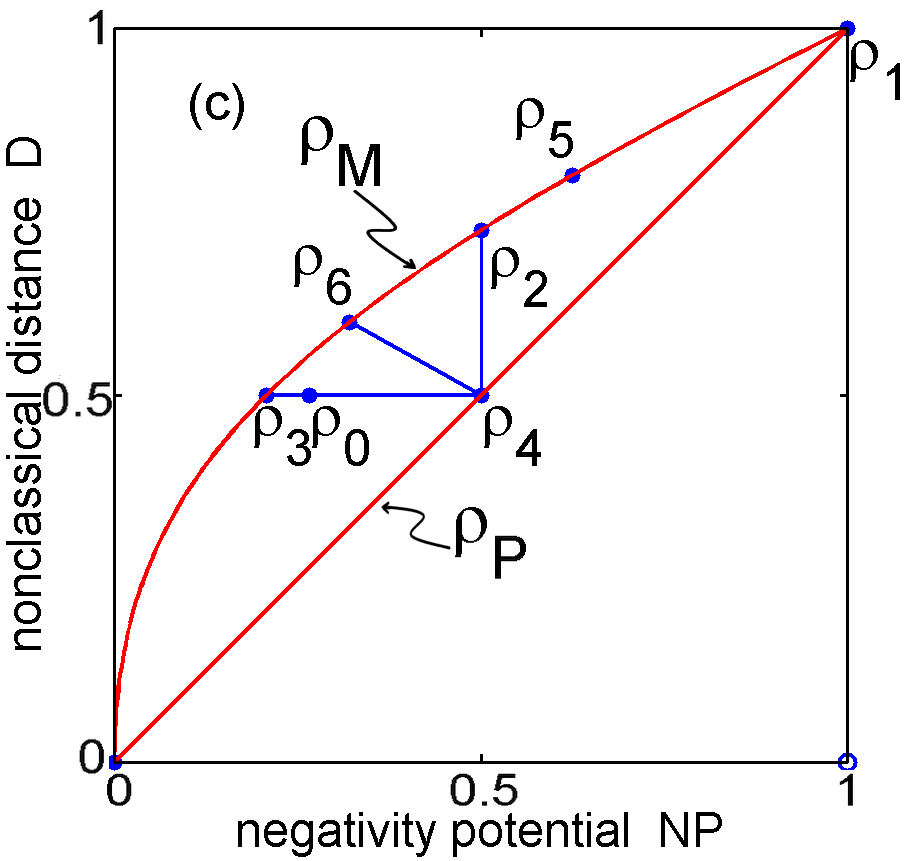}

\caption{(Color online) Particular single-qubit states $\rho_n$
defined explicitly in Table~II and plotted in analogy to Fig.~1.
As in Fig.~1, here the boundaries are given by pure states
$\rho_{\rm P}$, completely mixed states $\rho_{\rm M}$, together
with the partially mixed states $\rho_+$ and $\rho_{\rm opt}$.
There are no states corresponding exactly to the broken lines and
empty circles. Any inequality listed in Table~III can be satisfied
by properly choosing pairs of these states. This analysis
demonstrates the \emph{relativity of nonclassicality measures}.}
\end{figure*}
\end{widetext}

\begin{table} 
\caption{Definition of states $\rho_n$, also shown in Fig.~3, and
the analytical values of their four nonclassicality measures.
These states are chosen for discussion of the relativity of the
nonclassicality ordering of general states. }
\begin{tabular}{l c c c}
\hline state $\rho_n$ & $\tau(\rho)$   & $D(\rho)=\CP(\rho)$ &
$\NP(\rho)$ \\ [2pt]  \hline
$\rho_0=\rho(\frac12,\frac14)$ & $\frac47$ & $\frac12$ &
$\cos(\frac29 \pi)-\frac12$ \\ [2pt]
$\rho_1=\ket{1}\bra{1}$ & 1 & 1 & 1 \\ [2pt]
$\rho_2=\rho_{\rm M}[\frac12(\sqrt{6}-1)],$ &
$\frac12(\sqrt{6}-1)$ & $\frac12(\sqrt{6}-1)$ & $\frac12$ \\ [2pt]
$\rho_3=\rho_{\rm M}(\frac12)$ & $\frac12$ & $\frac12$ &
$\frac12(\sqrt{2}-1)$ \\ [2pt]
$\rho_4=\rho_{\rm P}(\frac12)$ & 1 & $\frac12$ & $\frac12$ \\
[2pt]
$\rho_5=\rho_{\rm M}(\frac45)$ & $\frac45$ & $\frac45$ &
$\frac{1}{5} \left(\sqrt{17}-1\right)$ \\ [2pt]
$\rho_6=\rho_{\rm M}({\frac35})$ & ${\frac35}$ & ${\frac35}$ &
$\frac{1}{5} \left(\sqrt{13}-2\right)$ \\ [2pt] \hline
\end{tabular}
\end{table}
\begin{table} 
\caption{Inequalities and examples of pairs of states
$(\rho_n,\rho_m)$ satisfying them. The states $\rho_n$ (with
$n=1,\ldots,6$) are defined in Table~II and are plotted in Fig.~3.
Some inequalities imply same orderings, and others involve
different orderings of single-qubit states by the nonclassical
measures: depth $\tau$, distance $D$, and negativity potential
NP.}
\begin{tabular}{l c c c}
\hline
\hline
1 \quad\quad  & $\tau(\rho_1)>\tau(\rho_2)$ & \quad and \quad~ &$D(\rho_1)>D(\rho_2)$\\
2 &$ \tau(\rho_1)=\tau(\rho_4)$ & \quad and \quad~ &$D(\rho_1)>D(\rho_4)$\\
3 & $\tau(\rho_4)>\tau(\rho_3)$ & \quad and \quad~ &$D(\rho_4)=D(\rho_3)$\\
4 & $\tau(\rho_2)<\tau(\rho_4)$ & \quad and \quad~ &$D(\rho_2)>D(\rho_4)$\\
\\
5 & $\tau(\rho_1)>\tau(\rho_2)$ & \quad and \quad~ &$\NP(\rho_1)>\NP(\rho_2)$\\
6 & $\tau(\rho_1)=\tau(\rho_4)$ & \quad and \quad~ &$\NP(\rho_1)>\NP(\rho_4)$\\
7 & $\tau(\rho_4)>\tau(\rho_2)$ & \quad and \quad~ &$\NP(\rho_4)=\NP(\rho_2)$\\
8 & $\tau(\rho_5)<\tau(\rho_4)$ & \quad and \quad~ &$\NP(\rho_5)>\NP(\rho_4)$\\
\\
9 &  $\NP(\rho_1)>\NP(\rho_2)$& \quad and \quad~ &$D(\rho_1)>D(\rho_2)$\\
10 & $\NP(\rho_2)=\NP(\rho_4)$& \quad and \quad~ &$D(\rho_2)>D(\rho_4)$\\
11 & $\NP(\rho_4)>\NP(\rho_3)$& \quad and \quad~ &$D(\rho_4)=D(\rho_3)$\\
12 & $\NP(\rho_6)<\NP(\rho_4)$& \quad and \quad~ &$D(\rho_6)>D(\rho_4)$ \\
\hline \hline
\end{tabular}
\end{table}
\subsection{Relativity of nonclassicality measures}

The nonclassicality measures can give different predictions not
only concerning the absolute values, but more importantly,
regarding the ordering of states. In other words, by comparing two
states we cannot usually judge which of them is more nonclassical.

It is somehow surprising that any pure state (different from the
vacuum) has the same maximum nonclassicality with respect to the
nonclassical depth, which is not the case for the other discussed
measures.

A natural conjecture concerning basic properties of good
nonclassicality measures can be formulated as follows: By
comparing the values of such measures for a pair of arbitrary
states $\rho'$ and $\rho''$, one can order them uniquely.
Specifically, they should have the same degree of nonclassicality
or one of them should be less nonclassical than the other
according to all good nonclassicality measures. For example, if
$\tau(\rho')<\tau(\rho'')$, then the same inequality should also
hold for other measures including the NP and $D$. However, one can
falsify this conjecture by recalling a deeper relation between
some nonclassicality and entanglement measures and by referring to
the works, where the relativity of entanglement measures has
already been demonstrated~\cite{Eisert99, Miran04c,Miran04d,
Miran04a,Miran04b}. Here detailed comparisons, shown in Table~III
and Fig.~3, give evidence for this relativity even for
nonclassicality measures, which are not directly related to
entanglement.

\section{Conclusions}

Various measures of the amount of nonclassicality have been
proposed with respect to the definition of nonclassicality based
on the nonpositivity of the $P$~function. Here we have applied the
following measures to quantify the nonclassicality of single-qubit
states: the Lee nonclassical depth $\tau$, the Hillery
nonclassicality distance $D$, and the entanglement potentials NP
and CP.

We have found analytical expressions for these measures for the
simplest nontrivial example of single-qubit photon-number states.
These formulas clearly show the relativity of ordering states with
nonclassicality measures, as summarized in Tables~I and~III. Only
the CP and nonclassical distance were found to be equivalent.

Further, we have found maximally and minimally nonclassical states
by comparing any two of these measures. Surprisingly, statistical
mixtures of states can be more nonclassical than their
superpositions. Indeed, mixed states are the most nonclassical if
one considers the nonclassicality distance for a given value of
either the nonclassical depth or of the NP in the whole range
$[0,1]$ of the abscissa, as well as the NP versus the nonclassical
depth $\tau$ such that $\tau\ge\tau_0$, where
$\tau_0=0.3154\ldots$, as shown in Fig.~1. However, there are
partially mixed states, which have the NP for a given value of
$\tau\in[0,\tau_0)$ slightly larger than for completely mixed
states, as shown in Figs. 1(b) and 2(a).

Both of our results, concerning (i) the relativity of ordering
states with nonclassicality measures and (ii) the nonclassicality
of mixed states exceeding that of superposition states, are a
consequence of the nonequivalence of some of the most popular
measures of nonclassicality, including the nonclassical depth,
nonclassical distance, and NP. There are also equivalent measures,
including the nonclassical distance, CP, and the potential for the
entanglement of formation, as given by Eq.~(\ref{EoF}). Clearly,
the above mentioned counterintuitive properties do not appear for
such equivalent measures.

We found that the nonclassical distance $D$, as defined for the
specific choice of the reference classical states, corresponds to
the CP for arbitrary single-qubit states. This result shows an
operational interpretation of this nonclassical distance as the
potential for the entanglement of formation.

The present analysis can be extended to similar comparative
studies of other quantitative measures of nonclassicality of
single-, two-, and multimode systems. In particular, one can focus
on the comparative approaches to quantify the nonclassicality of
correlations as listed in, e.g., Ref.~\cite{Nakano13}.

We believe that our study could further stimulate interest in the
nonclassicality measures applied to finite-dimensional systems in
finding their general properties including their operational
interpretations.

\acknowledgments The authors thank Micha\l{} Horodecki and Marcin
Paw\l{}owski for stimulating discussions. A.M. is supported by the
Polish National Science Centre under Grants No.
DEC-2011/03/B/ST2/01903 and DEC-2011/02/A/ST2/00305. K.B.
acknowledges support by the Foundation for Polish Science and the
Polish National Science Centre under Grant No.
DEC-2013/11/D/ST2/02638, the Operational Program Research and
Development for Innovations European Regional Development Fund
(Project No. CZ.1.05/2.1.00/03.0058), and the Operational Program
Education for Competitiveness European Social Fund (Project No.
CZ.1.07/2.3.00/20.0017) of the Ministry of Education, Youth and
Sports of the Czech Republic. A.P. thanks Department of Science
and Technology (DST), India for support provided through the DST
Project No. SR/S2/LOP-0012/2010. J.P.Jr acknowledges Project No.
LO1305 of the Ministry of Education, Youth and Sports of the Czech
Republic and Project No. P205/12/0382 of GA \v{C}R. F.N. is
partially supported by the RIKEN iTHES Project, MURI Center for
Dynamic Magneto-Optics, and a Grant-in-Aid for Scientific Research
(S).

\appendix

\section{Proofs for boundary states}

Here we prove that completely mixed states $\rho_{\rm M}$, pure
states $\rho_{\rm P}$, and partially mixed states, $\rho_{\rm
opt}$ and $\rho_+$, are the boundary (or extremal) states shown in
Figs.~1 and~2. Specifically:

(1) The upper bound in Fig.~1(a): As $\tau(\rho) \ge D(\rho)$
holds for any single-qubit $\rho$ as given in Eq.~(\ref{ineq1}),
then it is seen that this bound is reached by the completely mixed
states $\rho_{\rm M}$, for which it holds $\tau(\rho_{\rm M})=
D(\rho_{\rm M})$.

(2) The upper bound in Figs.~1(b) and Fig.~2(a) was obtained
numerically by maximizing a single-variable function of the NP,
given in Eq.~(\ref{NPgeneral}) with $|x|^2=p-p^2/\tau$, for a
given $\tau$. In particular, the completely mixed states
$\rho_{\rm M}$ correspond to this upper bound for $\tau>\tau_0$.
Indeed, $\rho_{\rm M}$ satisfy the Karush-Kuhn-Tucker (KKT)
conditions, as can be shown analogously to the method applied for
the two-qubit measures of entanglement and Bell
nonlocality~\cite{Miran08b,Horst13,Bartkiewicz13}. We note that
these KKT conditions correspond to a refined method of Lagrange
multipliers~\cite{Boyd04}.

(3) The upper and lower bounds in Fig.~1(c): The area in the
relation between the nonclassical distance (or, equivalently, CP)
and NP of \emph{arbitrary single-qubit} states is the same as the
area in the relation between the concurrence and negativity of
\emph{arbitrary two-qubit} states. As shown in
Ref.~\cite{Verstraete01}, the two-qubit pure states and the
Horodecki states are the extremal states for the concurrence
versus negativity, but these states can be generated from the pure
and mixed single-qubit states, respectively, as discussed in
Sec.~III.A. Thus, the pure and mixed single-qubit states are the
extremal states for the relation between the nonclassical distance
and NP.

(4) The lower and right-hand bounds in Figs.~1(a) and~1(b) are
implied from the property that all these measures have their
values in the range $[0,1]$.

This concludes our proofs of the boundary states.


\end{document}